\documentclass[a4paper,11pt]{article}
\usepackage{jheppub}
\usepackage{graphicx}
\usepackage{bm}
\usepackage{color}
\usepackage{xcolor}
\usepackage{footmisc}
\usepackage[utf8]{inputenc}
\usepackage{amsmath}
\usepackage{fancyhdr}
\usepackage{wasysym}
\usepackage{amssymb}
\usepackage{feynmp-auto}
\usepackage{hyperref}

\title{ The 3-3-1 Model: a natural framework for sub-MeV dark matter}

\author[a,b]{Vinicius Oliveira}\emailAdd{viniciuslbo@ua.pt}

\author[c]{D. Cogollo}\emailAdd{diegocogollo@df.ufcg.edu.br}

\author[d]{A. Doff}\emailAdd{agomes@utfpr.edu.br}

\author[e]{C. A. de S. Pires}\emailAdd{cpires@fisica.ufpb.br}

\affiliation[a]{Departamento de F\'{i}sica da Universidade de Aveiro, Campus de Santiago, 3810-183 Aveiro, Portugal}

\affiliation[b]{Laborat\'{o}rio de Instrumenta\c{c}\~{a}o e F\'{i}sica Experimental de Part\'{i}culas (LIP), Universidade do Minho, 4710-057 Braga, Portugal}

\affiliation[c]{Departamento de Física, Universidade Federal de Campina Grande,
Caixa Postal 10071, 58109-970, Campina Grande, PB, Brazil}

\affiliation[d]{Universidade Tecnologica Federal do Parana - UTFPR - DAFIS, R. Doutor Washington Subtil Chueire, 330 - Jardim Carvalho, 84017-220, Ponta Grossa, PR, Brazil}

\affiliation[e]{Departamento de F\'isica, Universidade Federal da Para\'iba, Caixa Postal 5008, 58051-970, Jo\~ao Pessoa, PB, Brazil}

\abstract{
We show that the $\mathrm{SU}(3)_C \times \mathrm{SU}(3)_L \times \mathrm{U}(1)_N$ model with right-handed neutrinos naturally accommodates a viable sub-MeV dark matter (DM) candidate realized as pseudo-Goldstone boson that acquires  tiny  mass  through gravitational effects. The observed relic abundance is obtained via freeze-in in a low-reheating temperature scenario, without requiring tiny couplings. The model operates at the TeV scale and remains testable at current and future collider experiments.}

\begin{document}

\maketitle

\section{Introduction}

The nature of dark matter remains one of the most compelling open questions in physics \cite{Planck:2018vyg}. While weakly interacting massive particles (WIMPs) have long been considered the leading candidates \cite{Bertone:2004pz}, the absence of experimental evidence has motivated the exploration of alternative scenarios \cite{Arcadi:2017kky}, including light dark matter in the sub-MeV regime.

Thermal sub-MeV dark matter candidates are typically overproduced in the early Universe if standard cosmological assumptions are adopted \cite{Drewes:2016upu}. This issue can be addressed by considering non-standard thermal histories \cite{Dror:2020jzy,Bezrukov:2009th,Nemevsek:2012cd,Dutra:2021lto,Lebedev:2024mbj}. A particularly well-motivated alternative is provided by scenarios with low reheating temperatures. In this case, the low reheating temperature, $T_R$, suppresses the DM production. As a result, DM is produced out of equilibrium without requiring extremely feeble interactions, in contrast to the standard freeze-in paradigm \cite{Hall:2009bx}. This regime, often referred to as \textit{strong freeze-in} \cite{Cosme:2023xpa,Koivunen:2024vhr,Arcadi:2024wwg}, allows for richer phenomenology and opens the possibility of observable signatures associated with the DM interactions. 

Low-reheating temperature scenarios are also well motivated from a theoretical perspective. During and immediately after inflation, particles are  produced through gravitational interactions \cite{Lebedev:2022cic,Koutroulis:2023fgp}, generating an irreducible dark matter background. In standard freeze-in setups, this contribution can \textit{contaminate} the predicted relic abundance. However, if reheating occurs at sufficiently low temperatures, the pre-existing dark matter density is efficiently diluted, justifying the assumption of a negligible initial abundance in freeze-in calculations.

Extensions of the Standard Model (SM) based on the gauge symmetry $\text{SU}(3)_C \times \text{SU}(3)_L \times \text{U}(1)_N$ \cite{Singer:1980sw,Montero:1992jk,Foot:1994ym}, commonly known as 3-3-1 models, offer a rich framework for physics beyond the SM. In particular, the version with right-handed neutrinos (331RHN) provides a natural setting for light DM candidates \cite{Dutra:2021lto,Dias:2005yh}. Depending on the realization of the scalar sector, the model can accommodate either sterile neutrinos or pseudo-Goldstone bosons as viable DM candidates.

In this work, we investigate the possibility that the 331RHN model accommodates a sub-MeV DM candidate realized as pseudo-Goldstone boson. This state emerges from the spontaneous breaking of a hidden global symmetry, U$(1)_X$, inherent to the model. Its mass is generated through higher-dimensional operators, such as those induced by gravitational effects. We show that, in contrast to previous approaches relying on non-standard thermal histories, the correct dark matter relic abundance can be achieved within a low-reheating temperature scenario.

Specifically, we analyze the scalar and gauge sectors of the model, identify the relevant interactions responsible for dark matter production, and compute the relic abundance both analytically and numerically.

Our results demonstrate that the 331RHN model provides a consistent and testable framework for sub-MeV DM, which can be realized at the TeV scale and remains accessible to current and future collider experiments.

\section{Key aspects of the model}
\subsection{Fermion Content}
In the 331RHN model, right-handed neutrinos are incorporated as the third component of the leptonic triplets,
\begin{equation}
f_{aL}= \begin{pmatrix}
\nu_{a}     \\
l_{a}       \\
\nu^{c}_{a} \\
\end{pmatrix}_{L} \sim (1,3,-1/3), \quad e_{aR}\sim (1,1,-1)\,,
\label{leptons}
\end{equation}
with $a=1,2,3$ referring to the three generations.

For the quark sector, one generation comes in the triplet and the other two compose an anti-triplet representation. Thus, there are three possibilities, as discussed in Ref.~\cite{Oliveira:2022vjo}. The most studied case is one where  
\begin{eqnarray}
&&Q_{iL} = \left (
\begin{array}{c}
d_{i} \\
-u_{i} \\
d^{\prime}_{i}
\end{array}
\right )_L\sim(3\,,\,\bar{3}\,,\,0)\,,u_{iR}\,\sim(3,1,2/3),\,\,\,\nonumber \\
&&\,\,d_{iR}\,\sim(3,1,-1/3)\,,\,\,\,\, d^{\prime}_{iR}\,\sim(3,1,-1/3),\nonumber \\
&&Q_{3L} = \left (
\begin{array}{c}
u_{3} \\
d_{3} \\
u^{\prime}_{3}
\end{array}
\right )_L\sim(3\,,\,3\,,\,1/3),u_{3R}\,\sim(3,1,2/3),\nonumber \\
&&\,\,d_{3R}\,\sim(3,1,-1/3)\,,\,u^{\prime}_{3R}\,\sim(3,1,2/3),
\label{quarks} 
\end{eqnarray}
with $i=1,2$. The primed quarks are the new quarks but with the usual electric charges. 

\subsection{Scalar sector}

In order to generate the correct particle masses, the model requires only three scalar
triplets, namely,
\begin{eqnarray}
\eta = \left (
\begin{array}{c}
\eta^0 \\
\eta^- \\
\eta^{\prime 0}
\end{array}
\right ),\quad \,\rho = \left (
\begin{array}{c}
\rho^+ \\
\rho^0 \\
\rho^{\prime +}
\end{array}
\right ),\,
\quad \chi = \left (
\begin{array}{c}
\chi^0 \\
\chi^{-} \\
\chi^{\prime 0}
\end{array}
\right ),
\label{SC} 
\end{eqnarray}
with $\eta$ and $\chi$ transforming as $(1\,,\,3\,,\,-1/3)$, and $\rho$ as $(1\,,\,3\,,\,2/3)$.

It is usual to impose the Lagrangian of the model to be symmetric by  $Z_2$ discrete symmetry with,
\begin{equation}
(\rho\,,\,\eta\,,\,e_{a_R}, u_{a_R}, d_{a_R}) \to -(\rho\,,\,\eta\,,\,e_{a_R}, u_{a_R}, d_{a_R}).
\label{Z2}
\end{equation}
The most general potential that preserves lepton number and is invariant under $Z_2$ is
\begin{align} 
V(\eta,\rho,\chi)&=\mu_\chi^2 \chi^{\dagger} \chi +\mu_\eta^2 \eta^{\dagger} \eta
+\mu_\rho^2 \rho^{\dagger} \rho+\lambda_1(\chi^{\dagger}\chi)^2 +\lambda_2(\eta^{\dagger}\eta)^2
+\lambda_3(\rho^{\dagger}\rho)^2 \nonumber \\
&+ \lambda_4(\chi^{\dagger}\chi)(\eta^{\dagger}\eta)
+\lambda_5(\chi^{\dagger}\chi)(\rho^{\dagger}\rho)+\lambda_6
(\eta^{\dagger}\eta)(\rho^{\dagger}\rho) \nonumber\\
&+\lambda_7(\chi^{\dagger}\eta)(\eta^{\dagger}\chi)
+\lambda_8(\chi^{\dagger}\rho)(\rho^{\dagger}\chi)+\lambda_9
(\eta^{\dagger}\rho)(\rho^{\dagger}\eta)   \nonumber\\
&-\left( \dfrac{f}{\sqrt{2}} \epsilon^{ijk} \eta_i \rho_j \chi_k +  \mathrm{H.c.}\right). 
\label{PPLN}
\end{align}

Concerning the spontaneous breaking of the symmetry, the most studied scenario corresponds to one where only the neutral scalars $\eta^0$, $\rho^0$, and $\chi^{\prime 0}$ acquire vacuum expectation values (VEVs). In contrast, here we consider an extend setup in which $\eta^0$, $\rho^0$, $\chi^{\prime 0}$, and $\eta^{\prime 0}$ all develop VEVs. In this framework, the neutral scalar fields are expanded around their respective VEVs in the usual way,
\begin{equation}
   \varphi \;\to\; \frac{1}{\sqrt{2}}\left( v_{\varphi} + R_{\varphi} + i I_{\varphi} \right)\,,
   \label{vevs}
\end{equation}
where $\varphi = \eta^0,\, \rho^0,\, \chi^{\prime 0},\, \eta^{\prime 0}$. In this case the set of minimal conditions are given by,
\begin{align}
   & \mu_\chi^2+\lambda_1 v_{\chi^{\prime}}^2+\frac{1}{2} \lambda_4 \left(v_\eta^2 + v_{\eta^{\prime}}^2\right)+ \frac{1}{2}\lambda_5 v_\rho^2 + \frac{1}{2}\lambda_7 v^2_{\eta^{\prime}}  - \frac{1}{4} f\frac{v_\eta v_\rho}{v_{\chi^{\prime}}}=0\,, \nonumber \\
   & \mu_\eta^2+\lambda_2\left( v_\eta^2+v_{\eta^{\prime}}^2\right)+\frac{1}{2} \lambda_4 v_{\chi^{\prime}}^2+ \frac{1}{2}\lambda_6 v_\rho^2   - \frac{1}{4} \frac{f v_\rho v_{\chi^{\prime}}}{v_\eta}=0\,, \nonumber \\
   & \mu_\rho^2+\lambda_3 v_\rho^2+\frac{1}{2} \lambda_5 v_{\chi^{\prime}}^2+ \frac{1}{2}\lambda_6 \left(v_\eta^2 +v^2_{\eta^{\prime}}\right)  - \frac{1}{4} f\frac{ v_\eta v_{\chi^{\prime}}}{v_\rho}=0\,, \nonumber \\
   & \mu_\eta^2+\lambda_2\left( v_\eta^2+v_{\eta^{\prime}}^2\right)+\frac{1}{2} \lambda_4 v_{\chi^{\prime}}^2+ \frac{1}{2}\lambda_6 v_\rho^2 + \frac{1}{2}\lambda_7 v^2_{\chi^{\prime}}=0\,.
\end{align}
From the second and fourth conditions, one obtains
\begin{equation}
    f = -\frac{\lambda_7\, v_{\chi^\prime} v_\eta}{v_\rho}\,.
\end{equation}

Expanding the scalar potential and applying the minimization conditions, we obtain the following mass matrix for the CP-even scalar sector in the basis $(R_\chi , R_{\eta^{\prime}} , R_{\chi^{\prime}} , R_\eta , R_\rho)$:
\begin{equation}
\hspace*{-3cm}
M^2_R =
\begin{pmatrix}
-\dfrac{\lambda_7 v^2_{\eta^{\prime}}}{4} 
& 0 
& \dfrac{\lambda_7 v_\eta v_{\eta^{\prime}}}{4} 
& \dfrac{\lambda_7 v_{\chi^{\prime}} v_{\eta^{\prime}}}{4}
& -\dfrac{\lambda_7 v_\eta v_{\chi^{\prime}} v_{\eta^{\prime}}}{4v_\rho}
\\
0 
& \lambda_2 v^2_{\eta^{\prime}} 
& \dfrac{1}{2}(\lambda_4 + \lambda_7)v_{\chi^{\prime}} v_{\eta^{\prime}}
& \lambda_2 v_\eta v_{\eta^{\prime}}
& -\dfrac{\lambda_7}{4} \dfrac{v_\eta v_{\chi^{\prime}} v_{\eta^{\prime}}}{v_\rho}
\\
\dfrac{\lambda_7 v_\eta v_{\eta^{\prime}}}{4}
& \dfrac{1}{2}(\lambda_4 + \lambda_7)v_{\chi^{\prime}} v_{\eta^{\prime}}
& \lambda_1 v^2_{\chi^{\prime}}
& \dfrac{1}{2}(\lambda_4 + \lambda_7) v_{\chi^{\prime}} v_\eta
& \dfrac{1}{2}\lambda_5 v_{\chi^{\prime}} v_\rho + \dfrac{\lambda_7}{4} \dfrac{v_{\chi^{\prime}} v^2_\eta}{v_\rho}
\\
\dfrac{\lambda_7 v_{\chi^{\prime}} v_{\eta^{\prime}}}{4}
& \lambda_2 v_\eta v_{\eta^{\prime}}
& \dfrac{1}{2}(\lambda_4 + \lambda_7) v_{\chi^{\prime}} v_\eta
& \lambda_2 v_\eta^2 - \dfrac{\lambda_7}{4} v^2_{\chi^{\prime}}
& \dfrac{1}{2}\lambda_6 v_\eta v_\rho + \dfrac{\lambda_7}{4} \dfrac{v^2_{\chi^{\prime}} v_\eta}{v_\rho}\\
-\dfrac{\lambda_7 v_\eta v_{\chi^{\prime}} v_{\eta^{\prime}}}{4v_\rho}
& \dfrac{1}{2}\lambda_6 v_\rho v_{\eta^{\prime}}
& \dfrac{1}{2}\lambda_5 v_{\chi^{\prime}} v_\rho + \dfrac{\lambda_7}{4} \dfrac{v_{\chi^{\prime}} v^2_\eta}{v_\rho}
& \dfrac{1}{2}\lambda_6 v_\eta v_\rho + \dfrac{\lambda_7}{4} \dfrac{v^2_{\chi^{\prime}} v_\eta}{v_\rho}
& \lambda_3 v^2_\rho - \dfrac{\lambda_7}{4} \dfrac{v^2_{\chi^{\prime}} v^2_\eta}{v^2_\rho}
\end{pmatrix}\,.
\label{mms}
\end{equation}

Upon diagonalizing this mass matrix, we obtain one CP-even scalar with vanishing mass, which corresponds to the Goldstone boson absorbed by the neutral gauge boson $U^0$. The remaining four CP-even scalars are massive. Among them, one state must reproduce the observed Higgs boson with mass $125~\text{GeV}$, which we denote by $h_1$. The masses of the other three scalars depend on the parameters of the scalar potential.

Observe in $M_R^2$ that the first two rows and columns  are proportional to $v_{\eta^{\prime}}$, which must be much smaller than the other VEVS, as will be shown below. Therefore, it is a good approximation to decouple these two degrees of freedom, $R_\chi$ and $R_{\eta^{\prime}}$, from the remaining ones. Under this assumption, we recover the $3 \times 3$ CP-even scalar mass matrix found \cite{Pinheiro:2022bcs}. Working in this decoupling limit, we identify,
\begin{equation}
    h_1 \approx \cos\varphi \, R_\eta + \sin\varphi \, R_\rho \,, 
    \qquad
    h_2 \approx -\sin\varphi \, R_\eta + \cos\varphi \, R_\rho \,,
    \label{ms}
\end{equation}
with  $h_1$ playing the role of  the SM-like Higgs boson. We assume, for simplicity, that $\varphi = \pi/4$. 

Concerning the CP-odd neutral scalars, the mass matrix in the basis 
$(I_{\eta^{\prime}}, I_\chi, I_{\chi^{\prime}}, I_\eta, I_\rho)$ 
takes the form
\begin{equation}
\hspace*{-3cm}
M^2_I =
\begin{pmatrix}
0 & 0 & 0 & 0 & 0 \\

0 & -\dfrac{\lambda_7 v^2_{\eta^{\prime}}}{4} 
&  \dfrac{\lambda_7 v_\eta v_{\eta^{\prime}}}{4} 
&  \dfrac{\lambda_7 v_{\eta^{\prime}} v_{\chi^{\prime}}}{4} 
& \dfrac{\lambda_7 v_\eta v_{\chi^{\prime}} v_{\eta^{\prime}}}{4v_\rho} \\

0 & \dfrac{\lambda_7 v_\eta v_{\eta^{\prime}}}{4}  
& - \dfrac{\lambda_7 v^2_\eta}{4} 
& -\dfrac{\lambda_7 v_\eta v_{\chi^{\prime}}}{4} 
& \dfrac{\lambda_7 v^2_\eta v_{\chi^{\prime}}}{4v_\rho}  \\

0 & \dfrac{\lambda_7 v_{\eta^{\prime}} v_{\chi^{\prime}}}{4} 
& -\dfrac{\lambda_7 v_\eta v_{\chi^{\prime}}}{4} 
& -\dfrac{\lambda_7 v^2_{\chi^{\prime}}}{4} 
& \dfrac{\lambda_7 v^2_{\chi^{\prime}} v_\eta}{4v_\rho} \\

0 &  \dfrac{\lambda_7 v_\eta v_{\chi^{\prime}} v_{\eta^{\prime}}}{4v_\rho} 
& \dfrac{\lambda_7 v^2_\eta v_{\chi^{\prime}}}{4v_\rho} 
& \dfrac{\lambda_7 v^2_{\chi^{\prime}} v_\eta}{4v_\rho} 
& \dfrac{\lambda_7 v^2_{\chi^{\prime}} v^2_\eta}{4v^2_\rho}
\end{pmatrix}\,.
\label{MDKd}
\end{equation}

From this matrix, we observe that $I_{\eta^{\prime}}$ is massless and decouples from the remaining pseudoscalar states. Upon diagonalizing the residual $4 \times 4$ mass matrix, we obtain three Goldstone bosons: one is absorbed by the $Z^{\prime}$ gauge boson, another by the $Z$ boson, and the third by $U^{0\dagger}$. 

The remaining physical state is a linear combination of $I_\eta$ and $I_\rho$, with a mass proportional to $v_{\chi^{\prime}}$. What is relevant for our purposes is that the pseudoscalar $I_{\eta^{\prime}}$ remains completely decoupled from the other states and appears as a pseudo-Goldstone boson, being massless at tree level.

\subsection{Gauge Boson Sector}

In the gauge boson sector, the original formulation of the 331RHN model reproduces  the standard gauge bosons, $W^{\pm}_\mu\,,\, Z_\mu\,,\, A_\mu$,  and introduces additional ones, $W^{\prime \pm}_\mu\,,\, U^{0}_\mu\,,\, U^{0 \dagger}_\mu\,,\, Z^{\prime}_\mu$ \cite{Long:1996rfd}. The presence of $v_{\eta^{\prime}}$ induces  mixing between the Standard Model particles and the new states.  In particular, the VEV $v_{\eta^{\prime}}$ generates mixing among the charged gauge bosons $W^+_\mu$ and $W^{\prime +}_\mu$, leading to the mass matrix
\begin{equation}
 \frac{g^2}{4}
\begin{pmatrix} 
v_\eta^2 + v_\rho^2 & v_{\eta} v_{\eta^{\prime}} \\
v_{\eta} v_{\eta^{\prime}} &  v_{\chi^{\prime}}^2 + v_{\eta^{\prime}}^2 + v_\rho^2 
\end{pmatrix},
\label{charged-mass}
\end{equation}
in the basis $( W^+_\mu ,\, W^{\prime +}_\mu)$. Diagonalizing this matrix, we obtain the mass eigenstates
\begin{equation}
\begin{pmatrix}
\hat W^+ \\
\hat W^{\prime +}
\end{pmatrix}
=
\begin{pmatrix} 
c_\theta & -s_\theta \\
s_\theta & c_\theta
\end{pmatrix}
\begin{pmatrix}
W^+_\mu \\
W^{\prime +}_\mu
\end{pmatrix},
\label{ccmixing}
\end{equation}
with corresponding eigenvalues
\begin{equation}
m^2_{\hat W^{+}} = \frac{g^2 v_{\mathrm{ew}}^2}{4}\left(1 - \frac{v_{\eta^{\prime}}^2}{v_{\mathrm{ew}}^2} + \frac{v_\rho^2}{v_{\chi^{\prime}}^2}\right)\,,
\qquad
m^2_{\hat W^{\prime +}} = \frac{g^2}{4}\left(v_{\chi^{\prime}}^2 + v_\rho^2 \right)\,,
\label{CCeigenvectors}
\end{equation}
where $v_{\mathrm{ew}}^2 = v_\eta^2 + v_\rho^2=246$ GeV. We define $c_\theta = \cos\theta$ and $s_\theta = \sin\theta$, with
\begin{equation}
\tan\theta \simeq \frac{v_{\eta^{\prime}} v_\eta}{v_{\chi^{\prime}}^2}\,.
\label{mgbcc}
\end{equation}

Concerning the neutral gauge bosons, a non-vanishing $v_{\eta^{\prime}}$ leads to significant modifications in this sector. To make this explicit, we observe that the mass matrix for the neutral gauge bosons in the basis $( W^3_\mu ,\, W^8_\mu,\, B_\mu ,\, W^4_\mu)$\footnote{The states $W^1_\mu\,,\, W^2_\mu\,,\, W^3_\mu\,,\, W^4_\mu\,,\,W^5_\mu\,,\,W^6_\mu\,,\,W^7_\mu\,,\,W^8_\mu\,,\,B_\mu$ are the symmetric ones associated to the $\mathrm{SU}(3)_L \times \mathrm{U}(1)_N$\cite{Long:1996rfd}} takes the form
\begin{equation}
\frac{g^2}{4}
\begin{pmatrix} 
v_{\mathrm{ew}}^2 
& \dfrac{v_\eta^2 - v_\rho^2}{\sqrt{3}} 
& -\dfrac{2}{3}(v_\eta^2 + 2v_\rho^2)t 
& v_{\eta^{\prime}} v_\eta
\\
\dfrac{v_\eta^2 - v_\rho^2}{\sqrt{3}} 
& \dfrac{1}{3}\left(v_{\mathrm{ew}}^2 + 4v_{\eta^{\prime}}^2 + 4v_{\chi^{\prime}}^2\right)  
& \dfrac{2}{3\sqrt{3}}\left(v_\eta^2 + 2v_\rho^2 + 2v_{\eta^{\prime}}^2 + 2v_{\chi^{\prime}}^2\right)t 
& -\dfrac{v_{\eta^{\prime}} v_\eta}{\sqrt{3}}
\\
-\dfrac{2}{3}(v_\eta^2 + 2v_\rho^2)t 
& \dfrac{2}{3\sqrt{3}}\left(v_\eta^2 + 2v_\rho^2 + 2v_{\eta^{\prime}}^2 + 2v_{\chi^{\prime}}^2\right)t  
& \dfrac{4}{9}\left( v_{\eta^{\prime}}^2 + v_{\chi^{\prime}}^2 + v_\eta^2 + 4v_\rho^2\right)t^2  
& -\dfrac{4 v_{\eta^{\prime}} v_\eta}{3}
\\
v_{\eta^{\prime}} v_\eta 
& -\dfrac{v_{\eta^{\prime}} v_\eta}{\sqrt{3}}    
& -\dfrac{4 v_{\eta^{\prime}} v_\eta}{3}      
& v_{\eta^{\prime}}^2 + v_{\chi^{\prime}}^2 + v_\eta^2
\end{pmatrix} \, ,
\end{equation}
 where $t=g_N/g$.

This matrix has a vanishing determinant, implying the existence of a massless eigenstate, which is identified as the photon. Its corresponding eigenvector is given by
\begin{equation}
    A_\mu = s_W W^3_\mu + c_W \left(-\frac{t_W}{\sqrt{3}} W^8_\mu + \sqrt{1 - \frac{t_W^2}{3}}\, B_\mu \right)\,,
\end{equation}
where $s_W = \sin\theta_W$, $c_W = \cos\theta_W$, and $t_W = \tan\theta_W$.

The set of orthonormal eigenvectors to $A_\mu$ is given by
\begin{align}
    Z_\mu &= c_W W^3_\mu - s_W \left(-\frac{t_W}{\sqrt{3}} W^8_\mu + \sqrt{1 - \frac{t_W^2}{3}}\, B_\mu \right)\,, \nonumber \\
    Z'_\mu &= \sqrt{1 - \frac{t_W^2}{3}}\, W^8_\mu + \frac{t_W}{\sqrt{3}} B_\mu \,.
\end{align}
These eigenstates are identified with the Standard Model neutral gauge boson $Z_\mu$ and the additional neutral gauge boson $Z^{\prime}_\mu$ present in the canonical 331RHN model. Proceeding with the mixing analysis, the photon $A_\mu$ decouples from the remaining neutral gauge bosons, which mix among themselves through the mass matrix in the basis $(Z,\, Z',\, W^4)$
\begin{equation}
\frac{g^2}{4c_W^2}
\begin{pmatrix} 
v_{\mathrm{ew}}^2 
& \beta\left(-v_\rho^2 + c_{2W} v_\eta^2\right) 
& c_W v_{\eta^{\prime}} v_\eta
\\
\beta\left(-v_\rho^2 + c_{2W} v_\eta^2\right) 
& \beta^2\left[4\left(v_{\eta^{\prime}}^2 + v_{\chi^{\prime}}^2\right)c_W^4 + v_\rho^2 + c_{2W} v_\eta^2\right]  
& -\beta c_W v_{\eta^{\prime}} v_\eta
\\
c_W v_{\eta^{\prime}} v_\eta 
& -\beta c_W v_{\eta^{\prime}} v_\eta  
& v_{\eta^{\prime}}^2 + v_{\chi^{\prime}}^2 + v_\eta^2
\end{pmatrix},
\label{neutral-mass2}
\end{equation}
where $\beta = \frac{1}{\sqrt{3 - 4s_W^2}}$ and  $c_{2W} = c_W^2 - s_W^2$. 

Let us now analyze this matrix. The mixing of $Z_\mu$ and $Z^{\prime}_\mu$ with $W^4_\mu$ is suppressed by a factor of order
\begin{equation}
\sim \frac{v_{\eta^{\prime}} v_\eta}{v_{\chi^{\prime}}^2}\,.
\end{equation}
As we will show below, $v_{\eta^{\prime}}$ is constrained to be much smaller than $v_\eta$, $v_\rho$, and $v_{\chi^{\prime}}$. Therefore, at least for the proposal of this work, the mixing of $Z_\mu$ and $Z^{\prime}_\mu$ with $W^4_\mu$ can be safely neglected. In doing this , $W^{4}_\mu$ couple with $W^5_\mu$ to form the neutral non-Hermitian gauge bosons $U^0$ and $U^{0\dagger}$ but now with mass,
\begin{equation}
M^2_{U^0} = \frac{g^2}{4}\left( v_{\eta^{\prime}}^2 + v_{\chi^{\prime}}^2 + v_\eta^2 \right)\,.
\end{equation}
Furthermore, the mixing between $Z_\mu$ and $Z^{\prime}_\mu$ is also strongly constrained, typically at the level of $\sim 10^{-4}$, see Refs.\cite{Long:1996rfd,Cogollo:2007qx} and can be neglected as well. Even under these approximations, the presence of a non-zero $v_{\eta^{\prime}}$ still modifies the mass terms. Considering all this, after diagonalization, we obtain
\begin{equation}
m^2_{Z} = \frac{g^2 v_{\mathrm{ew}}^2}{4c_W^2}\left(1 + \frac{4v_{\eta^{\prime}}^2 + v_{\mathrm{ew}}^2}{4 v_{\chi^{\prime}}^2}\right)\,,
\qquad
m^2_{Z'} = \frac{g^2 v_{\chi^{\prime}}^2 c_W^2}{3 - 4s_W^2}\,,
\end{equation}
Therefore, the presence of a non-zero $v_{\eta^{\prime}}$ modifies the canonical expressions for the masses of the $W^\pm_\mu$ and $Z_\mu$ gauge bosons, with important implications for the $\rho$ parameter, which now reads
\begin{equation}
    \rho = 1 + \frac{4v_{\chi^{\prime}}^2\left(v_\rho^2 v_{\mathrm{ew}}^2 + v_{\eta^{\prime}}^4\right) - 4v_{\eta^{\prime}}^2\left(v_{\mathrm{ew}}^2 v_\rho^2 + v_{\chi^{\prime}}^4\right) - 3v_{\eta^{\prime}}^2 v_{\mathrm{ew}}^2 v_{\chi^{\prime}}^2 - v_{\mathrm{ew}}^4\left(v_\rho^2 + v_{\chi^{\prime}}^2\right)}{4v^2 v_{\chi^{\prime}}^4}\,.
\end{equation}

The current experimental value for this parameter is \cite{ParticleDataGroup:2024cfk}
\begin{equation}
    \rho = 1.00031 \pm 0.00019\,.
\end{equation}
Taking $v_\eta = v_\rho = 174~\text{GeV}$ and $v_{\chi^{\prime}} = 10~\text{TeV}$, we obtain the upper bound
\begin{equation}
    v_{\eta^{\prime}} \lesssim 1.4~\text{GeV}\,.
\end{equation}

\subsection{Yukawa sector}

Under the discrete $Z_2$ symmetry, for the case of the third family transforming as triplet, the Yukawa interactions are composed by the terms
\begin{align}
-{\cal L}^Y &= f_{ij} \bar Q_{i_L}\chi^* d^{\prime}_{j_R} + f_{33} \bar Q_{3_L}\chi u^{\prime}_{3_R} + g_{ia}\bar Q_{i_L}\eta^* d_{a_R} + h_{3a} \bar Q_{3_L}\eta u_{a_R} \nonumber \\
&\quad + g_{3a}\bar Q_{3_L}\rho d_{a_R} + h_{ia}\bar Q_{i_L}\rho^* u_{a_R} + G_{l}\bar f_{l_L} \rho e_{l_R} + \text{H.c.} \,.
\label{yukawa1}
\end{align}
With this Yukawa structure, all fermions acquire masses, except the neutrinos.

Concerning neutrino masses, the $Z_2$  discrete symmetry above allows the following dimension-5 operators \cite{Dias:2005yh},
\begin{eqnarray}
    \frac{f^{\nu_L}}{\Lambda}(\bar f_L^C \eta^*) (\eta^{\dagger} f_L) +\frac{f^{\nu_R}}{\Lambda}(\bar f_L^C \chi^*) (\chi^{\dagger} f_L) \,.
    \label{effe}
\end{eqnarray}

The first operator generates Majorana masses for the standard neutrinos,
\begin{equation}
    m_{\nu_L}=f^{\nu_L}\frac{v^2_\eta}{\Lambda},
\end{equation}
while the second one generates Majorana masses for the right-handed neutrinos,
\begin{equation}
    m_{\nu_R}=f^{\nu_R}\frac{v^2_{\chi^{\prime}}}{\Lambda}.
\end{equation}
The $Z_2$ symmetry avoids the effective operator, $\frac{f^{\mathrm{mix}}}{\Lambda}(\bar f_L^C \chi^*) (\eta^{\dagger} f_L) $, that would generate  mixing  between right-handed and left-handed neutrinos. For $v_{\chi^{\prime}} \approx 10^4$GeV and $\Lambda \approx 10^{14}$ GeV the 331HRN model predicts  standard neutrinos with mass at sub-eV scale, and right-handed neutrinos with mass at sub-MeV regime.

\section{Dark matter}
\subsection{Goldstone as the dark matter candidate}
The original version of the  331RHN model does not contain a viable DM candidate, namely a neutral and stable (or sufficiently long-lived) particle. Under specific conditions, the lightest right-handed neutrino can play the role of a warm DM (WDM) candidate, as discussed in Ref.~\cite{Dutra:2021lto, Cabrera:2023rcy}. In this work, however, we relax these requirements and, instead, investigate the scalar sector as a potential source of a suitable DM candidate.

The presence of a pseudo Goldstone boson, namely $I_{\eta^{\prime}}$, signals that a global U$(1)$ symmetry is spontaneously broken by the vacuum expectation value $v_{\eta^{\prime}}$. The underlying origin of this global symmetry is not relevant for our purposes and will not be specified. From now on, we identify $I_{\eta^{\prime}} \equiv \phi$ as our dark matter candidate. 

In order to be phenomenologically viable, $\phi$ must acquire a mass. A straightforward possibility would be to introduce explicit soft-breaking terms of the global symmetry in the scalar potential. However, such terms typically induce mixing with other pseudoscalars, rendering $\phi$ unstable. Instead, we consider higher-dimensional operators induced by gravitational effects, which are expected to violate global symmetries. The leading contribution arises from the operator
\begin{equation}
    V(\eta,\rho,\chi) \supset \lambda \frac{\eta \rho \chi (\eta^\dagger \eta)}{M_{\mathrm{Pl}}} \,,
\end{equation}
which generates a mass for $\phi$ given by
\begin{equation}
    m^2_\phi = \lambda \frac{v_\eta v_\rho v_{\chi^{\prime}}}{2\sqrt{2}M_{\mathrm{Pl}}} \,.
\end{equation}
For $v_\eta = v_\rho = \frac{v_{\mathrm{ew}}}{\sqrt{2}}$, $v_{\chi^{\prime}} = 10~\mathrm{TeV}$ and $M_{\mathrm{Pl}} = 2.4 \times 10^{18}~\mathrm{GeV}$, we obtain
\begin{equation}
    m_\phi \simeq 30\,\sqrt{\lambda}\ \mathrm{keV}\,.
\end{equation}

We first need to ensure that $\phi$ is stable. Observe that $\phi$ is leptophobic. Being a singlet under the SM gauge symmetry, its Yukawa interactions necessarily involve 3-3-1 fermions. The Yukawa interactions in Eq.~\ref{yukawa1} allow for the following two-body decay channels $\phi \to \bar{q}_{\mathrm{SM}} + q_{\mathrm{331}}$ while the effective operators in Eq.~\ref{effe} allows the decay channel $\phi \to \bar{\nu}_R + \nu_L$, where $q_{\mathrm{331}}$ denotes the exotic quarks of the 3-3-1 model. However, since $m_{q_{\mathrm{331}}} \gg m_\phi$, the decay into quarks is kinematically forbidden. This leaves $\phi \to \bar{\nu}_R + \nu_L$ as the only possible decay channel. If $\phi$ is lighter than the right-handed neutrinos, this decay get also kinematically forbidden. Therefore, in what follows, we assume $m_\phi \ll m_{\nu_R}$, ensuring the stability of the dark matter candidate.

In summary, two factors ensure the stability of $\phi$. First,  the discrete $Z_2$  symmetry  that forbids  mixing among $\nu_R - \nu_L$ avoiding, in this way, the decay channel $\phi \to \nu_L +\nu_L$. Second, the supposition $m_\phi < m_{\nu_R}$ that forbids   $\phi \to \bar{\nu}_R + \nu_L$.

\subsection{Relic Abundance}

We now outline the cosmological history assumed throughout this work. At early times, the energy density of the Universe is dominated by the inflaton field. We consider a \textit{two-stage reheating} scenario, in which the inflaton predominantly decays into an intermediate sector before transferring its energy to SM particles (for more details, see Ref.\cite{Cosme:2024ndc}). We call reheating by the epoch at which the energy density of the SM radiation bath comes to dominate the total energy density of the Universe. An important consequence of this scenario is that the maximum temperature reached by the SM thermal bath is approximately the reheating temperature, $T_{\max} \simeq T_R$. As a result, the production of dark matter is controlled by $T_R$. In this work, however, we remain agnostic about the nature of the inflaton and the detailed dynamics governing its decay channels. Consequently, the reheating temperature, $T_R$, is taken as a free parameter of the model.

Our DM candidate interacts with the SM thermal bath via the scalars $h_1$ and $h_2$. The dominant production mechanism is
\begin{equation}
    \bar{f} f \to \phi \phi\,,
\end{equation}
where $f$ denotes SM fermions. Dark matter can also be produced through  $h_1 h_1 \to \phi \phi$. However, as will become clear below, this contribution is negligible in the present scenario, since the reheating temperatures considered here are well below the Higgs mass, leading to a strong Boltzmann suppression of its contribution. Interactions between $\phi$ and gauge bosons always involve one SM and one 3-3-1 gauge boson (i.e.\ $\phi$--$W_{\mathrm{SM}}$--$W_{\mathrm{331}}$), and are therefore subleading. Consequently, fermion annihilation processes provide the dominant contribution to the dark matter production in the temperature range of interest.

We are assuming that  among the new particles predicted by the 3-3-1 model, our dark matter candidate, $\phi$, is the lightest state. The interactions between the 3-3-1 states and the Standard Model particles are typically of the order of the electroweak coupling, or smaller. As a consequence, these particles do not  thermalize with the SM thermal bath in the temperature range considered in this work. This ensures that the dark matter production proceeds just through SM particles.

The reheating temperature should be $T_R \gtrsim 4$ MeV \cite{Hannestad:2004px}. Around the QCD phase transition, $T \sim 150~\mathrm{MeV}$, quarks become confined into hadrons. Consequently, for $T_R \lesssim 150~\mathrm{MeV}$, the hadronic production channels such as $\pi \pi$, $K \bar{K}$, \textit{etc.} should be taken into account \cite{Feiteira:2026qme}. In this work, however, we focus primarily on reheating temperatures above the QCD phase transition, $T_R \gtrsim 150~\mathrm{MeV}$. Whenever results for $T_R \lesssim 150~\mathrm{MeV}$ are presented, they should be understood as including only the fermionic contributions to the production rate. A more complete treatment, incorporating hadronic effects in this temperature regime, is beyond the scope of this work and will be left for future investigation.

The relevant interaction Lagrangian governing the production of dark matter is approximately given by
\begin{equation}
    \mathcal{L} \supset \frac{v_{\mathrm{ew}}}{2}\, \lambda_{+} \, h_1 \phi^2 
    + \frac{v_{\mathrm{ew}}}{2}\, \lambda_{-}\, h_2 \phi^2 
    + \frac{m_f}{v_{\mathrm{ew}}} h_1 \bar{f} f 
    + \frac{m_f}{v_{\mathrm{ew}}} h_2 \bar{f} f \,,
\end{equation}
where the couplings , $\lambda_{+}$ and $\lambda_{-}$, are defined as
\begin{equation}
    \lambda_{+} \equiv \frac{\sqrt{2}}{2}(2\lambda_2 + \lambda_6)\,, 
    \qquad 
    \lambda_{-} \equiv \frac{\sqrt{2}}{2}(2\lambda_2 - \lambda_6)\,.
\end{equation}
These interactions mediate the production of $\phi$ pairs from the thermal bath via $s$-channel scalar exchange.

This leads to the annihilation cross section
\begin{equation}
    \sigma(s)_{f\bar f \to \phi \phi} = 
    N_c \times \frac{1}{2} \times
    \frac{
    m_f^{2}\left(\lambda_{-} m_{h_1}^{2} + \lambda_{+} m_{h_2}^{2} - (\lambda_{-} + \lambda_{+})s\right)^{2}
    }{
    32 \pi \left(m_{h_1}^{2} - s\right)^{2}
    \left(m_{h_2}^{2} - s\right)^{2}
    }
    \sqrt{
    \frac{\left(s - 4 m_f^{2}\right)\left(s - 4 m_\phi^{2}\right)}{s^{2}}
    } \,,
\end{equation}
where $N_c = 3$ for quarks and $N_c = 1$ for leptons, and the factor $1/2$ accounts for the identical particles in the final state. It is important to note that in the limit $\lambda_{-} \to 0$ or $m_{h_2} \gg m_{h_1}$, the model reproduces the standard Higgs portal scenario \cite{Lebedev:2024mbj}.

The evolution of the DM number density $n_\phi$ is governed by the Boltzmann equation
\begin{equation}
\dot n_\phi + 3H n_\phi = 2 \Gamma(\mathrm{SM} + \mathrm{SM} \to \phi + \phi) - 2 \Gamma(\phi + \phi \to \mathrm{SM} + \mathrm{SM})\,,
\end{equation}
where the factor $2$ accounts for the two identical $\phi$ particles produced/annihilated per interaction. The Hubble parameter and entropy density for a radiation dominated Universe are given by
\begin{equation}
    H = \sqrt{\frac{\pi^2 g_e(T)}{90}}\frac{T^2}{M_\mathrm{Pl}}\,, \quad \quad  s_{\mathrm{SM}} = \frac{2\pi^2}{45} g_s(T)T^3\,,
\end{equation}
where $g_e$ and $g_s$ are the energetic and entropic degree of freedom, respectively. $M_\mathrm{Pl}$ is the reduced Planck mass. 

The reaction rate per unit volume for the process $f\bar{f} \to \phi\phi$ is given by
\begin{align}
\Gamma(f\bar{f} \to \phi\phi) &= \langle \sigma v_r \rangle n_f^{\text{(eq)} 2} \\
&= 2^2\, \frac{T}{32 \pi^4} \int_{\text{max}(4m_f^2,\, 4m_\phi^2)}^\infty d s \, 
\sigma_{f\bar{f} \to \phi\phi} (s - 4m_f^2) \sqrt{s} \, K_1 \left( \frac{\sqrt{s}}{T} \right)\,,
\end{align}
where the factor $2^2$ accounts for the two spin degrees of freedom of the initial-state fermions.

It is convenient to rewrite the Boltzmann equation in terms of the comoving yield $Y_\phi \equiv n_\phi/s_{\mathrm{SM}}$, which leads to
\begin{equation}
    \frac{dY_\phi}{dx} =  
    2 \sqrt{\frac{8\pi^2 M_\text{Pl}^2}{45}} 
    \frac{g_*^{1/2} m_\phi}{x^2}  
    \sum_{i}  \langle \sigma_{ f \bar{f} \to \phi\phi} v \rangle  
    Y^{\text{(eq)} 2}_f 
    \left[1 - \left(\frac{Y_\phi}{Y_\phi^\text{eq}} \right)^2 \right]\,,
\end{equation}
where $x \equiv m_\phi/T$. The effective number of relativistic degrees of freedom is defined as \cite{Gondolo:1990dk}
\begin{equation}
    g_*^{1/2} = \frac{g_s}{g_e^{1/2}} 
    \left( 1 + \frac{1}{3} \frac{T}{g_s} \frac{d g_s}{dT} \right)\,.
\end{equation}

The relic abundance of $\phi$ is then computed as
\begin{equation}
    \Omega_\phi h^2 = \frac{m_\phi Y_\phi^0 s_0 h^2}{\rho_{\mathrm{crit}}}\,,
\end{equation}
where $s_0 = 2 891.2$ cm$^{-3}$, $\rho_{\mathrm{crit}} = 1.053\times 10^{-5} h^2 \, \mathrm{GeV} \,  \mathrm{cm}^{-3}$ \cite{ParticleDataGroup:2024cfk}. The observed DM abundance, $\Omega h^2 = 0.12$ \cite{Planck:2018vyg}, corresponds to
\begin{equation}
    Y_\phi^0 \simeq \frac{4.4 \times 10^{-10}~\mathrm{GeV}}{m_\phi}\,.
\end{equation}

\subsubsection{Simple estimates of the DM relic abundance}

Since $\phi$ is produced non-thermally, the backreaction term can be safely neglected, namely $\Gamma(f  \bar f \to \phi  \phi) \gg \Gamma(\phi  \phi \to f  \bar f)$. The Boltzmann equation then simplifies to
\begin{align}    
    \frac{dY_\phi}{dx} &=  
    2 \sqrt{\frac{8\pi^2 M_\text{Pl}^2}{45}} 
    \frac{g_*^{1/2} m_\phi}{x^2}  
   \frac{\Gamma(f\bar f\to \phi \phi)}{s_\mathrm{SM}^2} \\ 
   &= \frac{27 M_\text{Pl}}{2 m_\phi^5 \pi^3} \int_{m_\phi/T_R}^\infty dx \, x^4 \Gamma(f\bar f\to \phi \phi)\,,
\end{align}
assuming $g_s= g_e = 10$.

We first consider the regime where the SM fermion is heavier than the DM particle ($m_f \gg m_\phi$) and the reheating temperature satisfies $T_R \ll m_f$. In this case, the minimal center-of-mass energy is $s_{\min} = 4m_f^2$, and the reaction rate reads
\begin{align}
\Gamma(f\bar{f} \to \phi\phi) = 2^2\, \frac{T}{32 \pi^4} \int_{4m_f^2}^\infty ds \, \sigma_{f\bar{f} \to \phi\phi} (s - 4m_f^2) \sqrt{s} \, K_1 \left( \frac{\sqrt{s}}{T} \right)\,,
\end{align}
where the cross section simplifies to
\begin{equation}
    \sigma(s)_{f\bar f \to \phi \phi} = 
    N_c \times \frac{1}{2} \times
    \frac{
    m_f^{2}\left(\lambda_{-} m_{h_1}^{2} + \lambda_{+} m_{h_2}^{2} \right)^{2}
    }{
    32 \pi \, m_{h_1}^{4} m_{h_2}^{4}
    }
    \sqrt{1 - \frac{4m_f^2}{s}}\,,
\end{equation}
where we have neglected the DM mass.

The integral appearing in $\Gamma$ can be approximated as
\begin{equation}
    \int_{4m_f^2}^\infty ds \, (s - 4 m_f^2)^{3/2} e^{-\sqrt{s}/T} 
    \simeq (4m_f T)^{5/2} e^{-2m_f/T} \times \frac{3}{4} \sqrt{\pi}\,.
\end{equation}

This leads to
\begin{equation}
\Gamma(f\bar{f} \to \phi\phi) =
N_c \times
\frac{
3 \, m_f^{4} \left(\lambda_{-} m_{h_1}^{2} + \lambda_{+} m_{h_2}^{2}\right)^{2}
}{
128 \, \pi^{4}\, m_{h_1}^{4} \, m_{h_2}^{4} 
} 
T^{4} e^{-\frac{2 m_f}{T}}\,.
\end{equation}

In the limit $m_{h_2} \gg m_{h_1}$, this expression simplifies to
\begin{equation}
\Gamma(f\bar{f} \to \phi\phi) =
N_c \times
\frac{
3 \, m_f^{4} \lambda_{+}^2
}{
128 \, \pi^{4}\, m_{h_1}^{4} 
} 
T^{4} e^{-\frac{2 m_f}{T}}\,,
\end{equation}
which reproduces the result obtained in Ref.~\cite{Lebedev:2024mbj}.

This yields the approximate abundance
\begin{equation}
Y_\phi \simeq 
N_c \times
\frac{
81 \, m_f^{3} \left(\lambda_{-} m_{h_1}^{2} + \lambda_{+} m_{h_2}^{2}\right)^{2} M_{\mathrm{Pl}}
}{
512 \, \pi^{7} m_{h_1}^{4} m_{h_2}^{4}
}
e^{-\frac{2 m_f}{T_R}}\,.
\end{equation}

As an illustration, the benchmark point $m_\phi = 100~\mathrm{keV}$, $m_f = m_\tau = 1.7~\mathrm{GeV}$, $m_{h_2} = 200~\mathrm{GeV}$, $T_R = 500~\mathrm{MeV}$, $\lambda_{+} = 0.1$, and $\lambda_{-} = -0.039$ reproduces the observed DM abundance while keeping $\phi$ out of equilibrium.

In the limit $m_{h_2} \gg m_{h_1}$, the yield simplifies to
\begin{equation}
Y_\phi \simeq 
N_c \times
\frac{
81 \, m_f^{3} \lambda_{+}^2 M_{\mathrm{Pl}}
}{
512 \, \pi^{7} m_{h_1}^{4}
}
e^{-\frac{2 m_f}{T_R}}\,.
\end{equation}

\vspace{0.3cm}

We now consider the opposite regime, where $m_f \gg m_\phi$ and $T_R \gg m_f$. In this case, the integrand is dominated by relativistic fermions, and we obtain
\begin{equation}
    \int_{4m_f^2}^\infty ds \, (s - 4m_f^2)^{3/2} K_1\left(\frac{\sqrt{s}}{T} \right) 
    \simeq \int_0^\infty ds \, s^{3/2} K_1\left(\frac{\sqrt{s}}{T} \right) = 32 \, T^5\,,
\end{equation}
which leads to
\begin{equation}
    \Gamma(f\bar f \to \phi \phi) \simeq 
    N_c \times 
    \frac{m_f^2 (m_{h_2}^2 \lambda_{+} + m_{h_1}^2 \lambda_{-})^2}{16 \pi^5 m_{h_1}^4 m_{h_2}^4} 
    T^6\,.
\end{equation}

The corresponding yield is
\begin{equation}
    Y_\phi \simeq 
    \frac{27 \, m_f^2 M_{\mathrm{Pl}} T_R (m_{h_2}^2 \lambda_{+} + m_{h_1}^2 \lambda_{-})^2}
    {32 \pi^8 m_{h_1}^4 m_{h_2}^4}\,.
\end{equation}

We observe that in the regime $m_f \gg T_R$, the abundance is exponentially suppressed, $Y_\phi \propto e^{-2m_f/T_R}$, whereas for $m_f \ll T_R$, it scales linearly with the reheating temperature, $Y_\phi \propto T_R$. Finally, a consistent computation of the relic abundance requires summing over all SM particles present in the thermal bath.  In what follows, we focus on the regime $T_R \lesssim 1~\mathrm{GeV}$ and solve the Boltzmann equation numerically including all SM fermions. 

\subsubsection{General case}

In order to obtain the DM abundance we need to solve numerically the Boltzmann equation
\begin{equation}
    \frac{dY_\phi}{dx} =   2 \sqrt{\frac{8\pi^2 M_\text{Pl}^2}{45}} \frac{g_*^{1/2} m_\phi}{x^2}  \sum_{f}  \langle \sigma_{ f \Bar{f} \to \phi\phi} v \rangle  Y^{\text{(eq)} 2}_f \times \left[1 - \left(\frac{Y_\phi}{Y_\phi^\text{eq}} \right)^2 \right]\,.
\end{equation}

\begin{figure}
    \centering
    \includegraphics[width=1.0\linewidth]{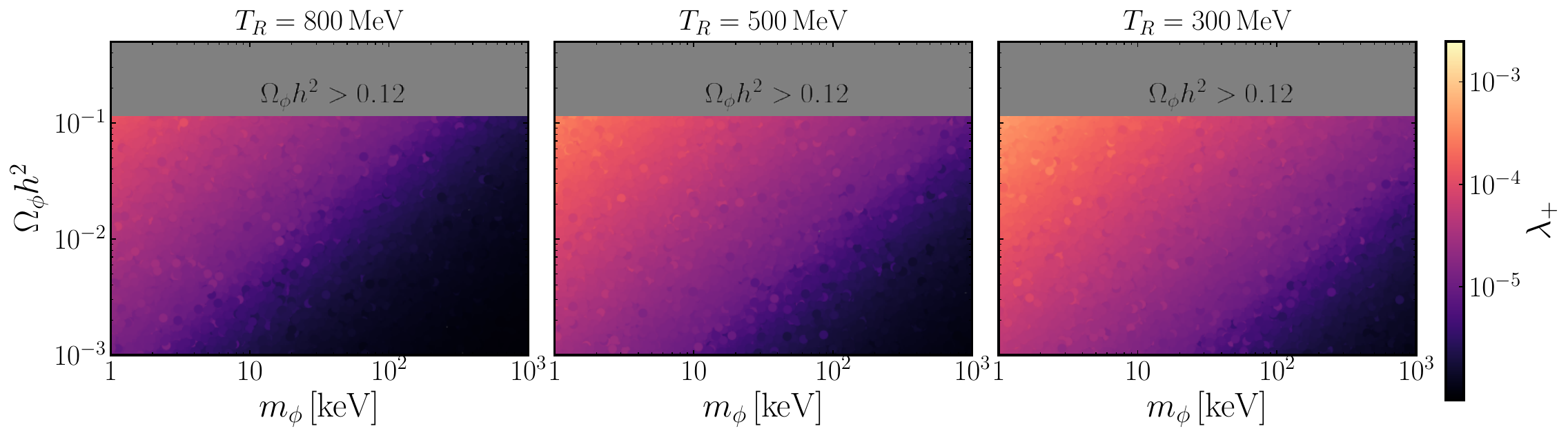}
    \caption{
Dark matter relic abundance in function of $m_\phi$ for different reheating temperatures, $T_R = 800,\,500,$ and $300~\mathrm{MeV}$, obtained by scanning over all relevant model parameters  ($ \lambda_{+}, \lambda_{-}, m_{h_2}$), with  $10^{-2} \geq \lambda_\pm \geq 10^{-7}$ and  $150~\mathrm{GeV} \leq m_{h_2} \leq 100~\mathrm{TeV}$. All displayed points satisfy current LHC constraints.  The gray region is excluded and corresponds to an overproduction of dark matter. The color coding represents the value of the coupling $\lambda_{+}$.
}
    \label{fig:full_parameter_space}
\end{figure}

Figure~\ref{fig:full_parameter_space} shows the results of our full numerical scan over the parameter space for three representative reheating temperatures, namely $T_R=800~\mathrm{MeV}$, $500~\mathrm{MeV}$, and $300~\mathrm{MeV}$. The figure displays the resulting dark matter relic abundance, $\Omega_\phi h^2$, as a function of the dark matter mass $m_\phi$, with the color code indicating the corresponding values of the coupling $\lambda_{+}$. The gray region denotes the portion of the parameter space where dark matter is overproduced, and is therefore excluded. In this scan, all relevant free parameters of the model were varied ($ \lambda_{+}, \lambda_{-}, m_{h_2}$), while retaining only those points consistent with current LHC bounds.

A clear correlation emerges between the relic abundance, the dark matter mass, and the coupling $\lambda_{+}$. As expected, larger values of $\lambda_{+}$ enhance the production rate of $\phi$, leading to larger values of $\Omega_\phi h^2$, whereas smaller couplings suppress the production and drive the relic abundance below the observed value. This behavior is visible in all three panels, where the abundance increases smoothly towards the upper-left region of the plots, while the lower-right region corresponds to progressively underabundant dark matter. 

Another important feature is the dependence on the reheating temperature.  By comparing the three panels, one observes that lower values of $T_R$ require larger couplings in order to reproduce the same relic abundance.  This behavior is a direct consequence of the reduced production efficiency in a \textit{colder} thermal bath: as the reheating temperature decreases, the DM production becomes  Boltzmann suppressed, and a stronger interaction strength is required to compensate for this suppression.

\subsection{Phase space distribution}
\label{sec:psd}

In this section, we compute the phase-space distribution (PSD) of the DM candidate, $\phi$, and assess its implications for structure formation. It has been recently pointed out \cite{Feiteira:2026qme} that freeze-in scenarios with low reheating temperatures generically lead to highly non-thermal momentum distributions, which cannot be captured by the commonly adopted $\alpha\beta\gamma$ parametrization \cite{Bae:2017dpt},
\begin{equation}
    f(q)\Big|_{\mathrm{high}\; T_R} \propto q^\alpha \, e^{\beta \, q^\gamma}\,,
\end{equation}
with constant coefficients $\alpha$, $\beta$, and $\gamma$.

The evolution of the distribution function $f_\phi(p,T)$ is governed by the Boltzmann equation in an expanding Universe,
\begin{equation}
\frac{\partial f_\phi}{\partial t} - H p \frac{\partial f_\phi}{\partial p} = \mathcal{C}[f_\phi] \,,
\end{equation}
where $\mathcal{C}[f_\phi]$ denotes the collision term. For the process $f\bar f \to \phi\phi$, it can be written as \cite{DEramo:2020gpr}
\begin{equation}\label{eq:Collision}
    \mathcal{C}[f_\phi] =  \frac{1}{16\pi^3}\frac{Te^{-E/T}}{p} \int_{4 m_f^2}^\infty ds \, \overline{|\mathcal{M}|^2} \sqrt{1- \frac{4m_f^2}{s}} \, e^{-\frac{s}{4pT}}\,,
\end{equation}
where we have neglected the DM mass, $m_\phi \ll m_f$. The squared matrix element, averaged over initial and final states and including the appropriate symmetry factors, is given by
\begin{equation}
    \overline{|\mathcal{M}|^2}_{f\bar f\to \phi\phi} = N_c\frac{ m_f^2 \left(s-4 m_f^2\right) \left(\lambda_{-} m_{h_1}^2+\lambda_{+} m_{h_2}^2-s (\lambda_{-}+\lambda_{+})\right)^2}{16 \left(m_{h_1}^2-s\right)^2 \left(m_{h_2}^2-s\right)^2}\,.
\end{equation}

The late-time distribution function can then be expressed as \cite{Feiteira:2026qme}
\begin{equation}\label{eq:psd}
f_\phi(q) = \frac{1}{q} \int_{T_0}^{T_R} \frac{dT}{T^2} \frac{\mathcal{C}(q,T)}{H}\,,
\end{equation}
where $q \equiv p/T$ and $T_0$ is the present temperature of the Universe. A key difference with respect to the standard freeze-in framework is that the integration is truncated at a finite reheating temperature $T_R$, rather than extending to $\infty$. This feature plays a crucial role in shaping the resulting non-thermal momentum distribution.

In our setup, the dominant production channels depend sensitively on the reheating temperature. For instance, for $T_R = 800~\mathrm{MeV}$, dark matter production is dominated by the process $c \bar c \to \phi \phi$, which accounts for approximately $65\%$ of the total relic abundance. A similar behavior is observed for $T_R = 500~\mathrm{MeV}$, where charm quark annihilations contribute about $81\%$ of the final abundance. For lower reheating temperatures, however, the charm contribution becomes exponentially suppressed, and lighter fermions progressively dominate the production.

To derive constraints from Lyman-$\alpha$ observations, we employ the area criterion, following Refs.~\cite{DEramo:2020gpr,Decant:2021mhj}. We first compute the PSD by solving Eq.~\eqref{eq:psd} numerically, using the full collision term in Eq.~\eqref{eq:Collision} and including all  production channels. The resulting distribution $f_\phi(q)$ is then tabulated and implemented in the Boltzmann solver CLASS \cite{blas2011cosmic}, which allows us to compute the linear matter power spectrum $P(k)$.

The suppression of structure is quantified by comparing the resulting power spectrum with that of the cold dark matter (CDM) case. This is done by defining the ratio $r(k)$ between the two spectra and introducing the associated area,
\begin{equation}
A_\phi = \int_{k_{\min}}^{k_{\max}} dk \, r(k)\,,
\end{equation}
where we adopt $k_{\min} = 0.5\,h/{\rm Mpc}$ and $k_{\max} = 20\,h/{\rm Mpc}$~\cite{Decant:2021mhj}. In the CDM limit, one recovers $A_{\rm CDM} = k_{\max} - k_{\min}$. It is then convenient to define the normalized area estimator,
\begin{equation}
\delta A_\phi = \frac{A_{\rm CDM} - A_\phi}{A_{\rm CDM}}\,,
\end{equation}
which quantifies the relative suppression of power.

In order to set bounds on the model, we compare $\delta A_\phi$ with the corresponding estimator computed for thermal WDM  benchmarks. We consider two representative limits derived from Lyman-$\alpha$ forest data \cite{Irsic:2017ixq,Garzilli:2019qki}, namely a conservative bound $m_{\rm WDM} > 1.9~\mathrm{keV}$ and a more stringent bound $m_{\rm WDM} > 5.3~\mathrm{keV}$. We first evaluate $\delta A_{\rm WDM}$ for these benchmark masses, and subsequently compute $\delta A_\phi$ for our model. Compatibility with Lyman-$\alpha$ observations requires that the suppression induced by our scenario does not exceed that of the corresponding benchmark, i.e. $\delta A_\phi \leq \delta A_{\rm WDM}$.

\begin{figure}
    \centering
    \includegraphics[width=1.\linewidth]{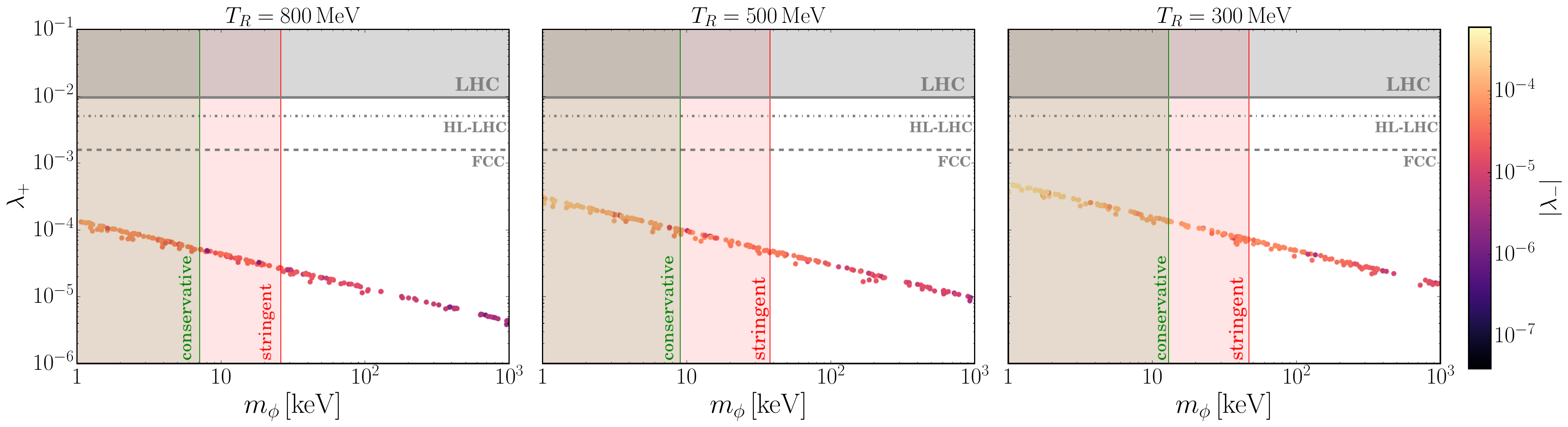}
    \caption{
Parameter space in the $(m_\phi, \lambda_+)$ plane for different reheating temperatures $T_R$, as indicated in each panel. The displayed points reproduce the observed dark matter relic abundance, $\Omega_\phi h^2 \simeq 0.12$. The color scale encodes the value of $|\lambda_-|$. The vertical shaded regions correspond to Lyman-$\alpha$ constraints, where the green (red) band represents the conservative (stringent) limit. Horizontal lines denote current and projected collider sensitivities from the LHC, HL-LHC and FCC experiments.
}
    \label{fig:placeholder}
\end{figure}

The resulting parameter space is shown in Figure \ref{fig:placeholder}. Each panel corresponds to a different reheating temperature $T_R$, illustrating how the cosmological history impacts both the production mechanism and the structure formation constraints. The displayed points reproduce the observed dark matter relic abundance, $\Omega_\phi h^2 \simeq 0.12$. The color coding reflects the magnitude of the coupling $|\lambda_-|$, highlighting the interplay between the two portal interactions. The vertical shaded regions indicate the exclusion limits from Lyman-$\alpha$ data, where the green (red) bands correspond to the conservative (stringent) WDM bounds. The shaded grey area is excluded by the LHC search for invisible Higgs decay ($\mathrm{BR}_\mathrm{inv} \lesssim 10\%$ \cite{ATLAS:2023tkt}), while the dash-dotted line represents sensitivity of the HL-LHC to
this mode. We take the HL-LHC benchmark goal to be $\mathrm{BR}_\mathrm{inv} \lesssim 3\%$ \cite{RivadeneiraBracho:2022sph}. While the dashed line represents the FCC sensitivity 
to be $\mathrm{BR}_\mathrm{inv} \lesssim 0.3\%$ \cite{FCC}.

\section{Conclusion}
In this work we developed a scenario that realizes  sub-MeV dark matter in the form of pseudo-Goldstone within the 331RHN model.  By  allowing the  two neutral scalars  of the triplet $\eta$ to  acquire VEVs, it gives rise to a pseudo‑Goldstone boson that serves as the dark matter candidate. Its tiny mass originates from gravitationally induced higher‑dimensional operators. Stability is ensured by assuming that the pseudo-Goldstone is lighter than the lightest right-handed neutrino. Its relic abundance is successfully reproduced through freeze‑in production in a low‑reheating temperature scenario, avoiding the need for extremely feeble couplings. This mechanism not only resolves the overproduction problem typical of light thermal dark matter but also leads to distinctive non‑thermal phase‑space distributions, with potential implications for structure formation. 

Unlike the scenario where right-handed neutrinos serve as sub-MeV dark matter candidates—forcing the 3-3-1 energy scale up to around $10^6$ GeV and thereby placing the 331RHN model beyond the reach of current and future colliders—the pseudo-Goldstone case keeps the 331RHN framework testable at the TeV scale. This means collider experiments such as the LHC, HL‑LHC, and FCC can probe both its scalar and gauge sectors, since, in general, the 331RHN model continues to be realized at accessible TeV energies.

 In summary, the 331RHN model stands out as a viable, predictive, and experimentally accessible theory for sub‑MeV dark matter, enriching both theoretical understanding and future observational prospects.

\section*{Acknowledgments}
V.O. is directly funded by FCT through the doctoral program grant with the reference PRT/BD/154629/2022 (\url{https://doi.org/10.54499/PRT/BD/154629/2022}). C.A.S.P  was supported by the CNPq research grants No. 311936/2021-0.

\bibliography{references}

\end{document}